\documentclass{IET-Conf-Paper}
\usepackage{graphicx}  
\usepackage{subfigure}    
\usepackage{enumerate}
\usepackage{algorithm}
\usepackage{algorithmic}
\usepackage{bm}
\counterwithout{figure}{section}
\counterwithout{equation}{section}
\counterwithout{table}{section}
\begin{document}

\title{TomoSAR-ALISTA: Efficient TomoSAR Imaging via Deep Unfolded Network}

\author{Muhan Wang\ad{1,2,3,4,5}, Zhe Zhang\ad{2,3,5}\corr, Yue Wang\ad{6}, Silin Gao\ad{1,4,5}, Xiaolan Qiu\ad{1,2,3,5}}

\address{
    \add{1}{Key Laboratory of Technology in Geo-spatial Information Processing and Application System, Chinese Academy of Sciences, Beijing 100190, China}
    \add{2}{Key Laboratory of Intelligent Aerospace Big Data Application Technology, Suzhou 215123, China}
    \add{3}{Suzhou Aerospace Information Research Institute, Suzhou 215123, China}
    \add{4}{School of Electronic, Electrical and Communication Engineering, University of Chinese Academy of Sciences, Beijing 100049, China}
    \add{5}{Aerospace Information Research Institute, Chinese Academy of Sciences, Beijing 100094, China}
    \add{6}{Electrical and Computer Engineering Department, George Mason University, Fairfax, VA 22030, USA}
    \email{zhangzhe01@aircas.ac.cn}}


\begin{abstract}
Synthetic aperture radar (SAR) tomography (TomoSAR) has attracted remarkable interest for its ability in achieving three-dimensional reconstruction along the elevation direction from multiple observations. In recent years, compressed sensing (CS) technique has been introduced into TomoSAR considering for its super-resolution ability with limited samples. Whereas, the CS-based methods suffer from several drawbacks, including weak noise resistance, high computational complexity and complex parameter fine-tuning. Among the different CS algorithms, iterative soft-thresholding algorithm (ISTA) is widely used as a robust reconstruction approach, however, the parameters in the ISTA algorithm are manually chosen, which usually requires a time-consuming fine-tuning process to achieve the best performance.  Aiming at efficient TomoSAR imaging,  a novel sparse unfolding network named analytic learned ISTA (ALISTA) is proposed towards the TomoSAR imaging problem in this paper, and the key parameters of ISTA are learned from training data via deep learning to avoid complex parameter fine-tuning and significantly relieves the training burden. In addition, experiments verify that it is feasible to use traditional CS algorithms as training labels, which provides a tangible supervised training method to achieve better 3D reconstruction performance even in the absence of labeled data in real applications. 
\end{abstract}

\maketitle

\section{Introduction}
SAR tomography (TomoSAR)~\cite{tomosar}, as an enabling RADAR imaging technique, has attracted remarkable interest in recent years for its ability in achieving three-dimensional resolution along the elevation direction from multiple observations. TomoSAR plays a significant role in broad applications, such as urban observation, forestry remote sensing and target identification~\cite{app1,app2}.

Traditionally, TomoSAR problems can be solved via spectrum estimation algorithms, which however usually experience poor performance under limited observations and low SNR circumstances. In typical sceanrios of TomoSAR applications, scatters are usually distributed sparsely along the elevation direction, and meanwhile only a few significant scatterers fall into a range-azimuth pixel. Thus, compressed sensing (CS)~\cite{CS} based methods are widely used to solve the TomoSAR inversion problem as the state-of-the-art approach.  

On the other hand, the CS-based methods suffer from several drawbacks, including weak noise resistance, high computational complexity and complex parameter fine-tuning. To overcome the above issues, different data-driven deep learning methods have been introduced into CS-based signal processing applications~\cite{DLCS1,DLCS2,DLCS3}. As a typical sparse signal processing application, deep learning methods have also been introduced and applied in TomoSAR imaging. For example, Budillon et al.~\cite{CNN} forumulated the TomoSAR inversion problem as a typical classification problem and utilized the neural networks to detect a single scatterer and estimate the corresponding elevation. ~\cite{CSDL} proposed an efficient line spectral estimation algorithm based on deep neural networks to tackle the TomoSAR inversion, which can distinguish overlaid scatterers and achieve diesirable estimation performance.  

Thanks to the enhanced interpretability, fast implementation, and high robusteness to model mistmatch, network unfolding technique has recently be proposed to develop fast neural network approximation for sparse optimization problems. Specifically, sparse reconstruction algorithms of CS can be represented by an unfolded deep network and solved via canonical training methods over a given dataset~\cite{LISTA}. Compared with traditional sparse reconstruction algorithms, the deep unfolded network is usually more robust, converges faster and does not require parameter fine-tuning. Kun Qian et al. developed a sparse unfolding network LISTA-CPSS~\cite{gammanet} to solve the TomoSAR problem under the name of $\gamma$-net. However, its computational cost is still relatively high, and it relies on a very large volume of dataset in the training stage, which are both remaining obstacles in practice of TomoSAR.

Aiming at efficient TomoSAR imaging, we focus on a holistic integration of advanced deep unfolding techinques with sparse signal processing algorithm. Along this line, a novel sparse unfolding network, called analytic learned ISTA (ALISTA), has been recently proposed as the unfolded network of the ISTA algorithm~\cite{ALISTA}, which is a popular sparse reconstruction algorithm and has been broadly utilized in TomoSAR problems~\cite{ISTASAR}. In ALISTA, the key parameters of ISTA are learned from training data via deep learning. Furthermore, the convergence speed of ALISTA is much faster than the standard ISTA. In this paper, as the first work of applying ALISTA to the TomoSAR problem, we customize ALISTA design to fit in the TomoSAR imaging model, which can avoid complex parameter fine-tuning and reduce computational complexity. In addition, an effective supervised training method is proposed for the real applications where the ground truth is not known as the training set.

The remainder of this paper is organized as follows. The problem statement of TomoSAR Imaging is first formulated in Section 2. Then, our TomoSAR imaging algorithm based on ALISTA unfolding network is proposed in Section 3. in Section 4 presents numerical results on image reconstruction by OMP, IHT, ISTA, and proposed ALISTA network on synthestic and real data, followed by conclusions in Section 5.

\section{TomoSAR Imaging Model}\label{section:model}
\begin{figure}[h]\vspace*{-1em}
    \centering
    \includegraphics[width=0.8\columnwidth]{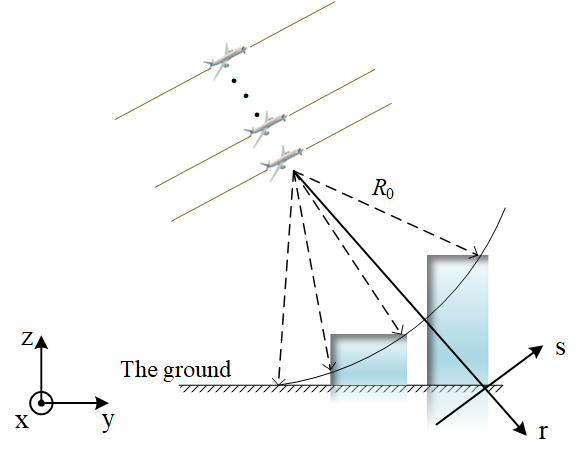}\vspace*{-1em}
    \caption{TomoSAR imaging model diagram.}
    \label{fig1}\vspace*{-1em}
\end{figure}

Tomographic SAR technique uses a stack of SAR images collected from different cross-track angles of the same observation area to reconstruct the scattering information along the elevation direction and achieve the three-dimensional resolving ability. The basic principle of SAR tomography is illustrated in Fig. \ref{fig1}.

Assume that the same target is observed N times from different viewing angles to obtain N single look complex  SAR images (SLCs). The values of the same pixel among N SLCs yields a sequence  . If the SLCs are perfectly aligned, each pixel value can be expressed as the integration of the upward scattering rate distribution along the elevation weighted by sinusoids as (1).

\begin{equation}\label{tomosar}
    y_n =\int{\gamma}(s) \exp \left( -j{2\pi}{\xi_n}s \right) \text{d}s\quad\left(n=1,2,...,N\right) 
\end{equation}

where $\xi_n=2b_n/\lambda r$ is the spatial frequency, $b_n$ is the baseline length the $n-$th channel and $\lambda , r$ is the wavelength and range. $\gamma \left(s\right)$ is the reflectivity function along the elevation. After discretizing the continuous elevation and considering the additive noise, Eq. (1) can be written in a matrix-vector form as (2).

\begin{equation}
\boldsymbol{y} = \mathbf{R}\bm{\gamma}+\varepsilon
\end{equation}

where $\mathbf{R}$ plays as a $N\times L$ sensing matrix

\begin{equation}
R_{nl} = \exp \left(-j2\pi \xi_ns_l\right)
\end{equation}

Now, TomoSAR imaging boils dwown to a signal recovery problem, where the goal is to obtain the reflectivity profile $\bm{\gamma}$ of each range-azimuth cell by solving (2).

\section{TomoSAR via ALISTA }
\subsection{ISTA and LISTA}
TomoSAR imaging usually becomes a sparse reconstruction problem~\cite{l1}, and can be solved via sparse reconstruction algorithms of CS such as OMP~\cite{OMP}, IHT~\cite{IHT}and ISTA~\cite{ISTA}. 

Among the different CS algorithms, Iterative soft-thresholding algorithm or ISTA is a widely used option, thanks to its robustness against noise and good reconstruction performance. In ISTA, the estimate of $\hat{\bm{\gamma}}$ is achieved under an iterative manner as

\begin{equation}
\bm{\gamma}_{k+1}=h_{(\alpha /L)}\left(\bm{\gamma}_k+\frac{1}{L}\mathbf{R}^\mathrm{H}(\bm{y}-\mathbf{R}\bm{\gamma}_k)\right)
\end{equation}
where $L$ is a parameter controlling the iteration step size and $L>\lambda_{\max}(\mathbf{R}^\mathrm{H}\mathbf{R})$, $\lambda_{\max}(\cdot)$ is an operation to calculate the maximum eigenvalue, $\theta=\alpha/L$ is the threshold parameter, and $h_{\theta}(\mathbf{X})$ is the soft-thresholding function defined as

\begin{equation}
h_{\theta}(\mathbf{X})=\mathrm{sign}(\mathbf{X})\max(|\mathbf{X}|-\theta, 0)
\end{equation}

However, the parameters $\alpha,L$ in the ISTA algorithm are manually chosen, which usually requires a time-consuming fine-tuning process to achieve the best performance. Furthermore, these parameters are not adaptive i.e., they are fixed from one scene to another.

As mentioned in~\cite{LISTA}, to let a neural network layer mimic the operation of signal processing algorithm, (4)  can be rewritten in a way of neuron’s activities:

\begin{equation}
\hat{\bm{\gamma}_{k+1}}=h_{\theta_{k+1}}(\mathbf{W}_1^{k+1}\bm{y}+\mathbf{W}_2^{k+1}\hat{\bm{\gamma_k}})
\end{equation}

where $\mathbf{W}_1^{k+1}=\frac{1}{L}\mathbf{R}^\mathrm{H}~$ and $~\mathbf{W}_2^{k+1}=1-\frac{1}{L}\mathbf{R}^\mathrm{H}\mathbf{R}$.\\

If we treat the soft-thresholding function $h_{\theta}(\cdot)$ as the activation function, then (7) is in the same form as the k+1th layer of a recurrent neural network (RNN). Inspired by the connection between ISTA and RNN, we can unfold the RNN into a deep network to leverage the benefits of deep learning, namely Learned ISTA (LISTA) as proposed in~\cite{LISTA}. LISTA is a sparse unfolded network that uses the learnability of deep learning to train the parameters from data: matrix $\mathbf{W}_1^{k},\mathbf{W}_2^{k}$ and threshold $\theta_{k}$, where k represents the number of layers in the deep network.

\subsection{ ALISTA Formulation for TomoSAR}

Although LISTA enables us to learn parameters from data autonomously, the weight matrices $\mathbf{W}_1$ and $\mathbf{W}_2$ are usually huge in real SAR signal processing applications. Hence, we need a vast dataset to train the enormous amount of parameters and the training stage is temporally and spatially complex. 

To solve  such challenges, we propose to customize Analytic LISTA (ALISTA) for TomoSAR, where both weight matrices are computed as the solution of a data-free convex optimization problem, while only the step size and threshold parameters are determined by data-driven learning. It significantly simplifies the training stage in both spatial and temporal domains. Here, the data-free optimization of the weight matrices is conducted based on coherence minimization.~\cite{ALISTA} indicates that ALISTA retains the benefits of LISTA in terms of optimal linear convergence rate and acheives a performance comparable to LISTA.

Instead of training $\mathbf{W}_1,\mathbf{W}_2$, in ALISTA $\mathbf{W}$ is pre-determined by solving the following convex optimization problem:

\begin{equation}
\mathbf{W} = \mathop{\arg \min}\limits_{\mathbf{W}\in \mathbb{C}^{M\times N}}\|\mathbf{W}_T\mathbf{R}\|_F^2\ \ \text{s.t.}(\mathbf{W}_{:.i}\mathbf{R}_{:,i},i=1,...,n)
\end{equation}

which plays as a mutual coherence minimizer between $\mathbf{W}$ and $\mathbf{R}$. This is motivated by the tenet in CS that a dictionary with smaller coherence possesses better sparse recovery performance. 

Then, let $\mathbf{W}^k=\mu^k\mathbf{W}$, (7) can be rewritten as the following:
\begin{equation}
\hat{\bm{\gamma}}_{k+1}=h_{\theta_{k}}\{\hat{\bm{\gamma}}_k+\eta_k\mathbf{W}^T(\bm{y}-\mathbf{R}\hat{\bm{\gamma}}_k)\}
\end{equation}
where only two scalar parameters  $\{\theta_k,\eta_k\}\in \mathbb{R}$  are learned from end to end. 

Thus, we can significantly reduce the number of training parameters ( from LISTA: $\mathcal{O}(KM^2+K+MN)$ down to ALISTA: $\mathcal{O}(K)$ ) and system complexity, which reduce the training burden and required training dataset size. The basic workflow of the proposed algorithm is shown in Algorithm 1.
\begin{algorithm}[htb]\vspace{0.2em} 
	\caption{TomoSAR-ALISTA }
	\label{alg:Framwork}
	\begin{algorithmic}
		\STATE \textbf{Obtain training data};\\
		\STATE \quad\textbf{Generate/Acquire} reflectivity profile $\boldsymbol{\gamma}$ as labels;\\
		\STATE \quad\textbf{Acquire} TomoSAR observation signal $\bm{y}$;\\
		\STATE \quad\textbf{Finish} The generation of training data $\{(\bm{y}_i,\bm{\gamma}_i)\}_{i=1}^{M\times N}$\vspace{0.5em}\\

		\STATE \textbf{Training of ALISTA}
		\STATE \quad\textbf{Generate} mapping matrix $\mathbf{R}$ with\vspace{0.5em}\\
		\STATE \quad\quad$\mathbf{R}_{nl}=\exp(-\frac{j4\pi b_ns_l}{\lambda r})$
		\vspace{0.5em}
		\STATE \quad\textbf{Precomputing $\mathbf{W}$} by solving a convex optimization problem\vspace{0.5em}\\
		\STATE \quad\quad $\mathbf{W} = \mathop{\arg \min}\limits_{\mathbf{W}\in \mathbb{C}^{M\times N}}\|\mathbf{W}_T\mathbf{R}\|_F^2\ \ \text{s.t.}(\mathbf{W}_{:.i}\mathbf{R}_{:,i},i=1,...,n)$\vspace{0.5em}\\
		
		\STATE \quad\textbf{The loss function} of ALISTA over the training data $\{(\bm{y}_i,\bm{\gamma}_i)\}_{i=1}^{M\times N}$ is defined as the mean square error loss: \vspace{0.5em}\\
		\STATE \quad\quad $\mathop{\min}\limits_{\Psi} L(\boldsymbol{\Psi})=\frac{1}{MN}\sum\limits_{i=1}^{MN}\|\hat{\boldsymbol{\gamma}}(\boldsymbol{\Psi} ,\mathbf{y})-\boldsymbol{\gamma}\|_2^2$\vspace{0.5em}\\
		\STATE\quad\quad $\hat{\bm{\gamma}}_{k+1}=h_{\theta_{k}}\{\hat{\bm{\gamma}}_k+\eta_k\mathbf{W}^T(\bm{y}-\mathbf{R}\hat{\bm{\gamma}}_k)$\vspace{0.5em}\\
		\STATE \quad\quad where $\bm{\Psi}=[\theta,\eta]$\\\vspace{0.5em}
		\STATE \quad\textbf{Complete} the training, obtain the trained model and optimization parameters
	\end{algorithmic}
\end{algorithm}

\section{Experiment results}
\subsection{Simulations}
\begin{figure}[h]\vspace*{-1em}
	\centering
	\includegraphics[width=0.75\columnwidth]{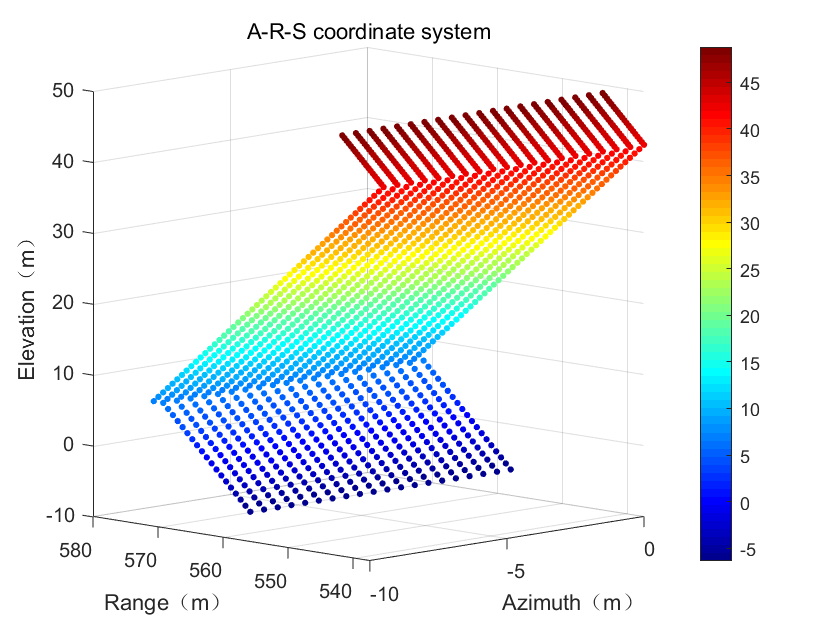}
	\caption{Simulated 3D building }
	\label{sim3D}\vspace*{-1em}
\end{figure}

A three-dimensional building structure is generated as the simulation target scene shown in Fig. \ref{sim3D}. And the simulation settings are given in Table 1.

\begin{table}[h]
    \caption{Simulation parameters}\vspace*{-1em}
    \setlength\tabcolsep{17pt}
    \begin{center}
        \begin{tabular}{ll}\toprule
            Parameters                                  & Values      \\\midrule
            Number of channels                     	    & 8           \\
            Wavelength (m) 						     	& 0.003125 \\
            Carrier frequency (GHz)                     & 5.5     \\
            Baseline interval (m)					    & 0.1 \\
            Average look angle ($^o$)			    	& 45  \\
            Distance of adjacent array elements (m)     & 0.1  \\
            Slant range of the middle trajectory (m)    & 400 \\
            SNR (dB)                                    & 20        \\\botrule
        \end{tabular}
        \label{tab1}
    \end{center}\vspace*{-2em}
\end{table}

\begin{figure}[h]\vspace*{-1em}
	\centering
	\includegraphics[width=0.75\columnwidth]{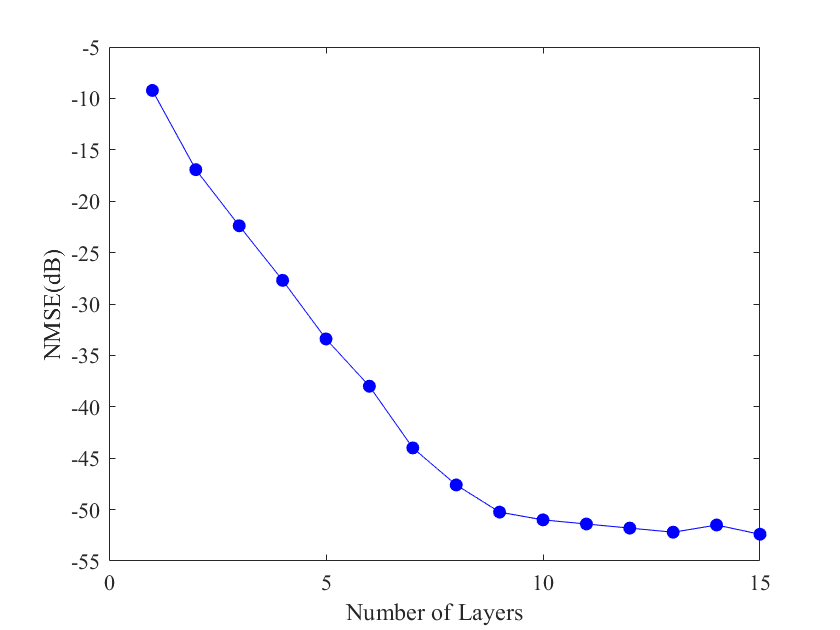}
	\caption{The performance of ALISTA w.r.t the number of layers }
	\label{LayerNum}\vspace*{-1em}
\end{figure}

We generate the training dataset in two ways. Method 1 uses the ground-truth value in simulation as the annotated label. Method 2 is to select some scattering points with good recovery accuracy and high SNR condition after reconstruction by traditional CS methods as data labels, considering that we usually lack ground truth in real applications. The training was carried out based on Python 3.7 using Tensorflow 1.12.0. In the training procedure, we gradually increase the number of layers from 1 to 15 to determine an optimal network structure. Fig. \ref{LayerNum} illustrates the performance of ALISTA with the different number of layers. It can be seen that the NMSE first decreases rapidly, and the benefit becomes marginal as the layer number reaches 10. The increment of layer number leads to a heavier computational burden. Therefore, ALISTA network employed in this article contains just ten layers. 

\begin{figure}[h!]\vspace*{-1em}
	\centering
	\subfigure[]{\includegraphics[width=0.3\textwidth]{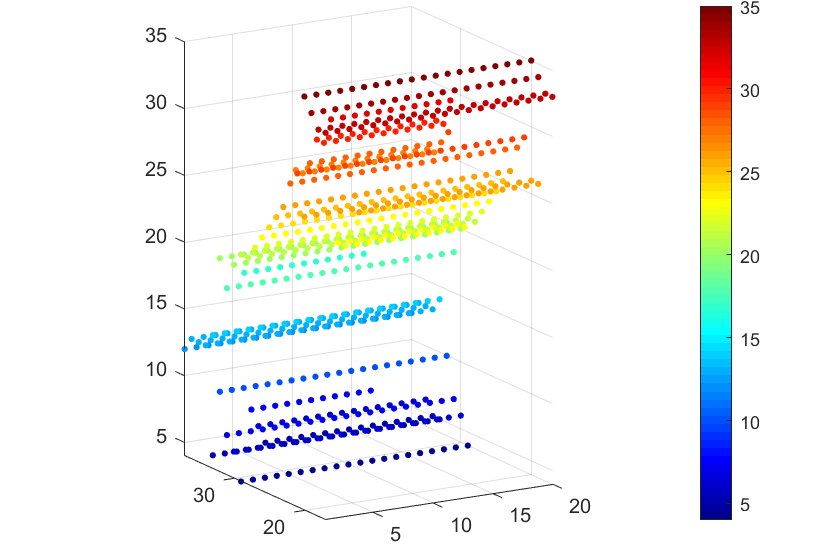}}\vspace*{-0.5em}
	\subfigure[]{\includegraphics[width=0.3\textwidth]{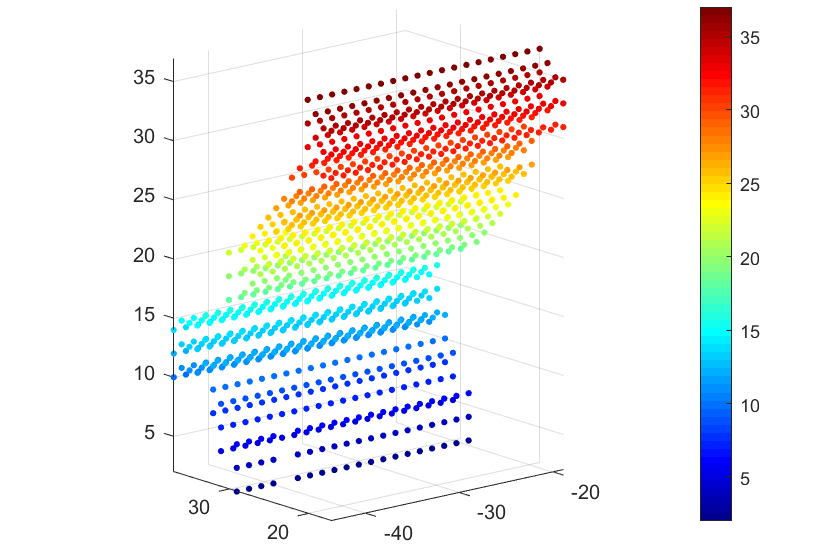}}\vspace*{-0.5em}
	\subfigure[]{\includegraphics[width=0.3\textwidth]{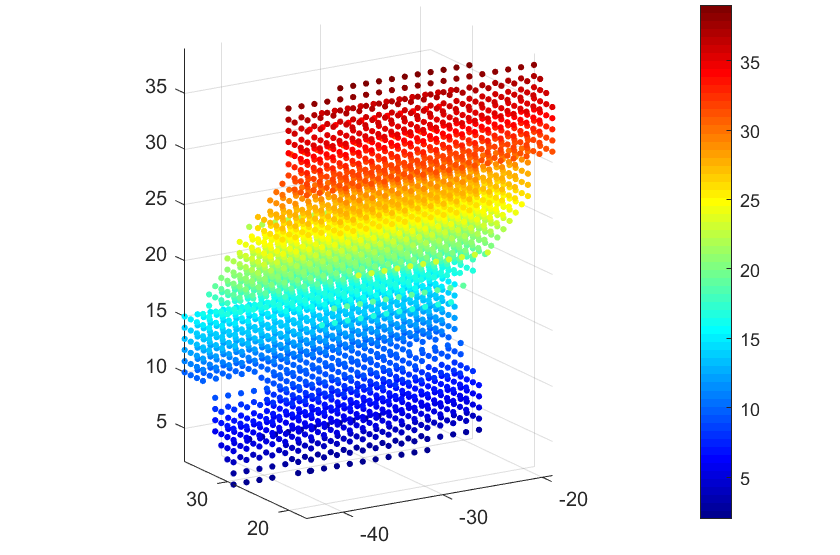}}\vspace*{-0.5em}
	\caption{Traditional results of 3D simulated building reconstruction:(a) OMP with sparsity K=2, (b) IHT with sparsity K=3 (c) traditional ISTA. }
	\label{simresult1}\vspace*{-1em}
\end{figure}

\begin{figure}[h!]
	\centering
	\subfigure[]{\includegraphics[width=0.3\textwidth]{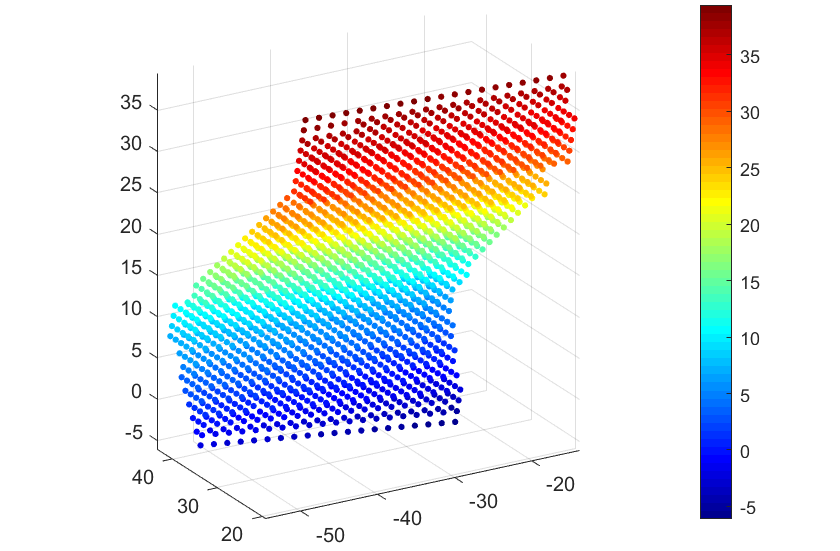}}\vspace*{-0.5em}
	\subfigure[]{\includegraphics[width=0.3\textwidth]{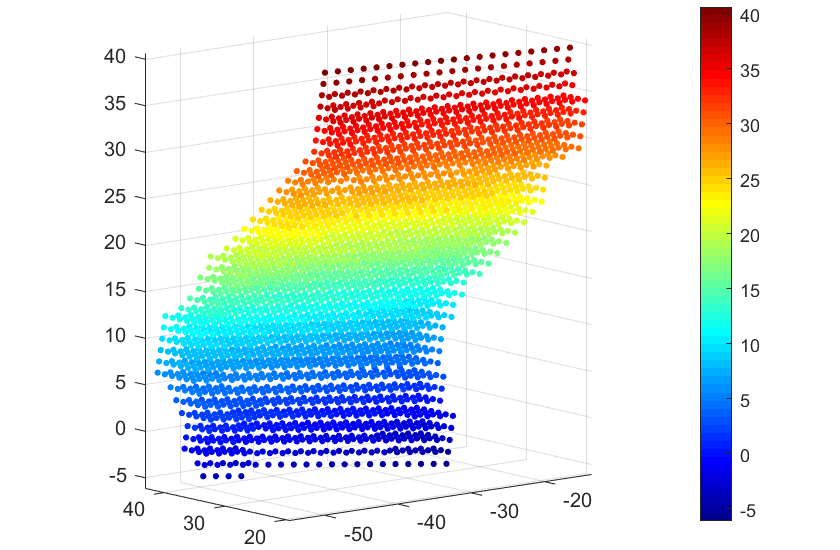}}\vspace*{-0.5em}
	\caption{TomoSAR-ALISTA results of 3D simulated building reconstruction:(a) TomoSAR-ALISTA trained from the ground truth label, (b) TomoSAR-ALISTA trained from the IHT label. }
	\label{simresult2}\vspace*{-2em}
\end{figure}

The training result in terms of three-dimensional point clouds is given in Fig. \ref{simresult1}.\ref{simresult2}. Reconstruction results of canonical OMP and ISTA algorithms are also provided as the benchmark. The trained step-size parameter $\eta$ was 0.02696917, and the threshold parameter $\theta$ was 0.01076145. 

We can see that both versions of ALSITA reconstruction results significantly outperform all three CS-based methods. Furthermore, the ALSITA result trained from the IHT label has a very similar visual effect as that trained from ground truth, revealing the possibility of utilizing IHT to generate annotated labels in real applications when we lack the ground truth. We also use NMSE to evaluate the accuracy of reconstruction results, which is shown in Table 2. The running time of OMP, IHT, traditional ISTA and ALISTA methods are also given in Table 3.

\begin{table}[h]\vspace*{-1em}
	\caption{NMSE of different algorithms}\vspace*{-1em}
	\setlength\tabcolsep{1pt}
	\begin{center}
		\begin{tabular}{llllll}\toprule
			Algorithm  & OMP  & IHT  & ISTA & ALISTA-GT & ALISTA-IHT  \\\midrule
			NMSE (-dB) &8.9594 &17.4396 &14.2579 &68.9500 &50.6792                 \\\botrule
		\end{tabular}
		\label{tab3}
	\end{center}\vspace*{-2em}
\end{table}

\begin{table}[h]\vspace*{-1em}
	\caption{The running time of different algorithms}\vspace*{-1em}
	\setlength\tabcolsep{8pt}
	\begin{center}
		\begin{tabular}{llllll}\toprule
			Algorithm  & OMP  & IHT  & ISTA & ALISTA  \\\midrule
			Time cost (s) & 0.25819  &0.24026 &1.57813 &0.26302        \\\botrule
		\end{tabular}
		\label{tab3}
	\end{center}
\end{table}

\vspace{-1em}
\subsection{Real Data Experiment}
\begin{figure}[h]
	\centering
	\subfigure[]{\includegraphics[width=0.2\textwidth]{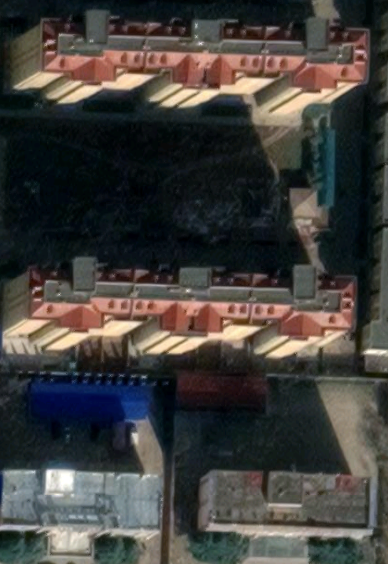}}
	\subfigure[]{\includegraphics[width=0.2\textwidth]{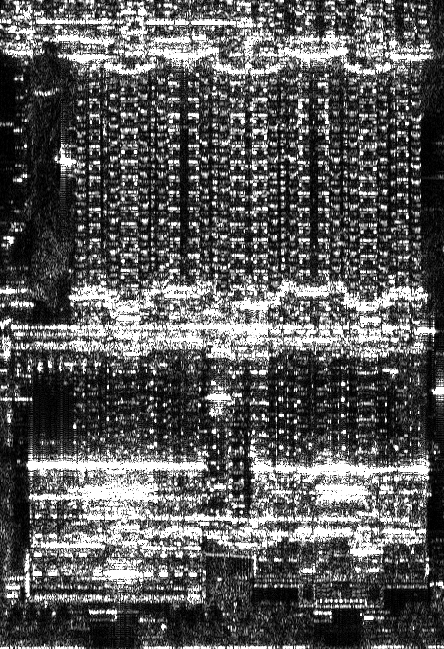}}
	\caption{Real data of Yuncheng area. (a) Optical image visualized in Google-Earth, (b) 2D SAR image.}
	\label{real}\vspace*{-1em}
\end{figure}

This section adopted real SAR data from the SARMV3D1.0 dataset~\cite{dataset}. It is an airborne array interferometric SAR system and the data is obtained from an urban community in Wanrong County, Yuncheng city, Shanxi Province, by Aerospace Information Research Institute, Chinese Academy of Sciences (AIRCAS). The array InSAR system has 8 channels. Optical and SAR images of the imaging area are shown in Fig. \ref{real}, in which the imaging targets are residential buildings with 14 floors. The height of each floor is around 3.5m, so the height of the building is approximately 52.5 meters. 

\subsubsection{Experiment setup} parameters and training label

TomoSAR experiment and system configuration is briefly enumerated in Table 4.

\begin{table}[h]
	\caption{Real data parameters}\vspace*{-1em}
	\setlength\tabcolsep{17pt}
	\begin{center}
		\begin{tabular}{ll}\toprule
			Parameters                                  & Values      \\\midrule
			Number of channels                     	    & 8           \\
			Wavelength (m) 						     	& 0.02105 \\
			Carrier frequency (GHz)                     & 14.25     \\
			Bandwidth (MHz)                             & 500   \\
			Baseline interval (m)					    & 0.0832 \\
			Average look angle ($^o$)			    	& 45  \\
			Distance of adjacent array elements (m)     & 0.1  \\
			Slant range of the middle trajectory (m)    & 959.0361 \\
			Waveband                                    & Ku        \\\botrule
		\end{tabular}
		\label{tab1}
	\end{center}\vspace*{-2em}
\end{table}

We currently lack precise ground truth of the imaging area at hand. Based on the simulation observations, it is feasible to select some imaging results of conventional CS methods with good reconstruction as the labeled data. Therefore, we adopt some carefully selected IHT reconstruction results as the training set.

\subsubsection{TomoSAR 3D Image Reconstruction Results}result and analysis

Traditional CS algorithms and the proposed ALISTA network are utilized, and the 3D cloud points are given as follows:

\begin{figure}[]\vspace*{-1em}
	\centering
	\subfigure[]{\includegraphics[width=0.35\textwidth]{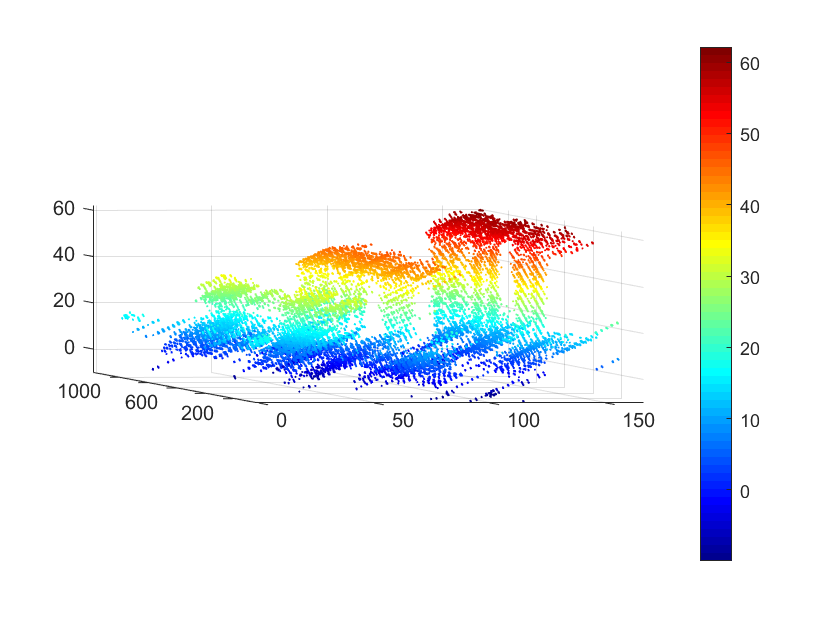}}\vspace*{-0.5em}
	\subfigure[]{\includegraphics[width=0.35\textwidth]{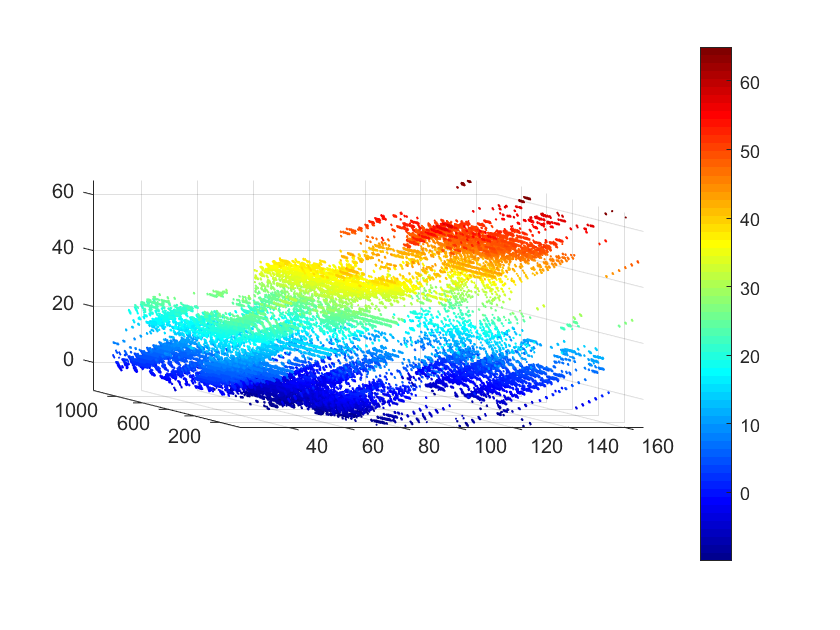}}\vspace*{-0.5em}
	\subfigure[]{\includegraphics[width=0.35\textwidth]{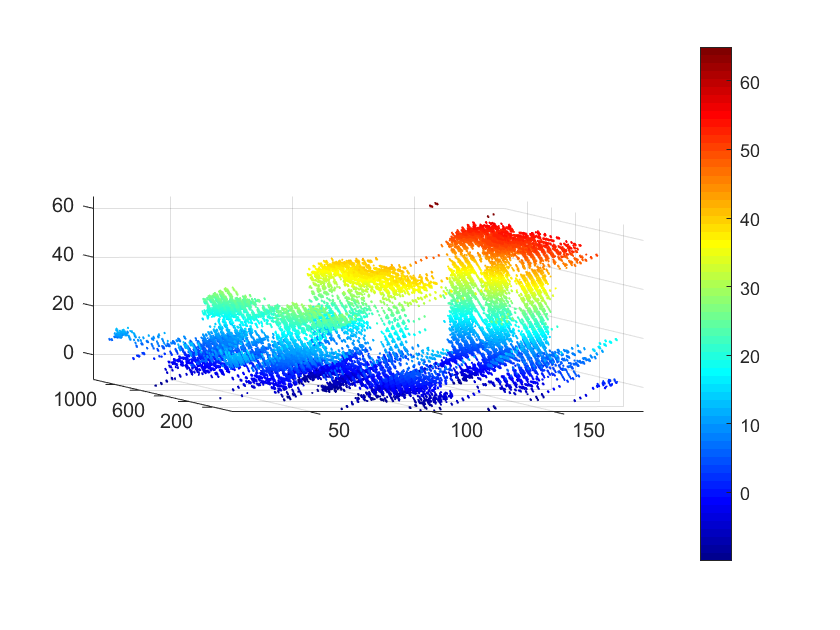}}\vspace*{-0.5em}
	\caption{Reconstructed and color-coded elevation of real data: (a) IHT, (b) IST, (c) Proposed ALISTA }
	\label{realresult}\vspace*{-1em}
\end{figure}

Processing with MATLAB R2018b and CPU intel i7-1165G7, the time cost of 3D image reconstruction is illustrated in Table 5.

\begin{table}[h]\vspace*{-1em}
	\caption{The running time of ISTA and propoed ALISTA}\vspace*{-1em}
	\setlength\tabcolsep{19pt}
	\begin{center}
		\begin{tabular}{lll}\toprule
			Algorithm    & ISTA & ALISTA  \\\midrule
			Time cost (s)  &2918.950596 &444.421808        \\\botrule
		\end{tabular}
		\label{tab5}
	\end{center}\vspace*{-2em}
\end{table}

Consistent with the simulated results on synthetic data, the reconstruction results of the proposed ALISTA Network are visually better than that of traditional CS algorithms on real data. Under the same experimental conditions and algorithm parameters, the time cost of traditional ISTA is approximately 6.6 times slower than the proposed ALISTA. Height of the reconstruction results is very close to 52.5m, which is consistent with the ground truth.

In summary, based on 3D image reconstruction results and analysis of simulated building and real data, we can conclude that the proposed method has higher reconstruction accuracy, better convergence rate, and higher algorithm efficiency than traditional CS algorithms under the same experimental conditions in TomoSAR applications.

\section{Conclusion}

In this paper, an efficient TomoSAR imaging method based on sparse unfolded network namely ALISTA is proposed. Compared with LISTA, ALISTA network reduces the parameter complexity of training from $\mathcal{O}(KM^2+K+MN)$ down to $\mathcal{O}(K)$, which significantly relieves the training burden in practice. At the same time, we also compare the network training effects by using different schemes in training set labels. The experiments show that it is feasible to use reconstruction results of traditional CS algorithms as training labels, which provides a tangible supervised training method for achieving better 3D reconstruction performance even in the absence of labelled data in real applications.

The future work includes: further improved ALISTA network; introducing the adaptive threshold method; exploring more domain knowledge in the iterative process of threshold shrinkage to achieve faster convergence and higher reconstruction accuracy; and using LiDAR/optical data as the training ground truth.

\section{Acknowledgements}

This work is supported by the NSFC grant \#62022082, 61991421, 61991420 and AIRCAS grant ``Structural sparsity signal high performance adaptive sensing theory and its applications in microwave imaging''.

\end{document}